\PassOptionsToPackage{table}{xcolor}
\documentclass{Interspeech}

\interspeechcameraready

\usepackage{algorithm}
\usepackage{amsmath,amsfonts,bm}
\usepackage{algpseudocode}
\usepackage{subfigure}
\usepackage{subcaption}

\renewcommand{\algorithmicensure}{\textbf{Output:}}

\renewcommand{\algorithmicensure}

\usepackage{multirow}%

\usepackage{xcolor}

\title{Regularized Federated Learning for Privacy-Preserving Dysarthric and Elderly Speech Recognition}

\author[affiliation={1}, equalcontribution]{Tao}{Zhong}
\author[affiliation={2}, equalcontribution] {Mengzhe}{Geng}
\author[affiliation={1}]{Shujie} {Hu}
\author[affiliation={1}]{Guinan}{Li}
\author[affiliation={1}]{Xunying}{Liu}

\affiliation{}{The Chinese University of Hong Kong}{China}
\affiliation{}{National Research Council Canada}{Canada}
\email{tzhong@se.cuhk.edu.hk, Mengzhe.Geng@nrc-cnrc.gc.ca, sjhu@se.cuhk.edu.hk, gnli@se.cuhk.edu.hk, xyliu@se.cuhk.edu.hk}


\usepackage{comment}

\usepackage{cite}
\newcommand{\tim}[1]{{\textcolor{black}{#1}}}
\newcommand{\review}[1]{{\textcolor{black}{#1}}}
\newcommand{\timreview}[1]{{\textcolor{black}{#1}}}
\usepackage{tabularx} 

\newcolumntype{Y}{>{\centering\arraybackslash}X}

\usepackage{pifont}
\usepackage{amssymb}

\newcommand{\xmark}{\ding{55}}

\usepackage[table]{xcolor} 

\begin{document}
\bstctlcite{IEEEexample:BSTcontrol}

\maketitle
\begin{abstract}

\tim{Accurate recognition of dysarthric and elderly speech remains challenging to date. While privacy concerns have driven a shift from centralized approaches to federated learning (FL) to ensure data confidentiality, this further exacerbates the challenges of data scarcity, imbalanced data distribution and speaker heterogeneity. To this end, this paper conducts a systematic investigation of regularized FL techniques for privacy-preserving dysarthric and elderly speech recognition, addressing different levels of the FL process by 1) parameter-based, 2) embedding-based and 3) novel loss-based regularization. Experiments on the benchmark UASpeech dysarthric and DementiaBank Pitt elderly speech corpora suggest that regularized FL systems consistently outperform the baseline FedAvg system by statistically significant WER reductions of up to 0.55\% absolute (2.13\% relative). Further increasing communication frequency to one exchange per batch approaches centralized training performance.}


\end{abstract}

\section{Introduction}
\label{sec:intro}

\tim{While automatic speech recognition (ASR) technologies targeting normal speech have advanced rapidly over the past decades~\cite{dong2018speech,gulati20_interspeech}, accurate speech recognition in the healthcare domain, particularly for dysarthric and elderly speakers, remains highly challenging to date~\cite{ye2021development,liu2021recent,wang22k_interspeech,geng2022speaker,yue2022acoustic,baskar2022speaker,10584335,jin2023personalized,wang2024enhancing,hsu2024cluster,wang2025phone}.}
\tim{Dysarthric and elderly speech introduces fundamental challenges for current deep learning-based ASR systems primarily targeting non-aged healthy users, including: \textbf{1) substantial mismatch} between such data and normal voices; \textbf{2) data scarcity} due to the difficulty of data collection, often limited by speakers' mobility issues \cite{10584335,liu2021recent}; and \textbf{3) large speaker heterogeneity} compounding accent, gender, and speech impairments or aging-induced neurocognitive decline \cite{kodrasi2020spectro}.}
\tim{Given the high priority of privacy in healthcare, there has been a growing shift towards decentralized training over centralized approaches, as it mitigates the risk of privacy leakage by limiting the exposure of sensitive data~\cite{williamson2024balancing}.}
\tim{Federated learning (FL)~\cite{pmlr-v54-mcmahan17a} proves to be an effective method for addressing data privacy concerns, as it enables collaborative model training across decentralized datasets without exchanging raw data.}
\tim{In recent years, there has been increasing attention on applying FL to speech-related tasks for the normal, healthy population, such as keyword spotting~\cite{leroy2019federated,hard2020training}, speaker verification~\cite{Granqvist2020,zhang2024stealthy}, speech emotion recognition~\cite{latif2020federated,feng2022semi}, and automatic speech recognition~\cite{dimitriadis2020federated,guliani2021training,yu2021federated,cui2021federated,nandury2021cross,gao2022end,azam2023importance,azam2023federated,pelikan2023federated,kan2024parameter,du2024communication}.}
\tim{Most prior work in FL-based ASR exclusively targets healthy adult speakers and adopts the Federated Averaging (FedAvg)~\cite{pmlr-v54-mcmahan17a} strategy to aggregate locally trained client models, where the parameters of the individual client are weighted by the proportion of its data samples relative to the total training samples.}
\tim{To constrain local clients from deviating too far from the global model, Federated Averaging with Diversity Scaling (FedAvg-DS) is proposed in\cite{nandury2021cross}, while FedProx~\cite{MLSYS2020_1f5fe839} is investigated in~\cite{pelikan2023federated} where a proximal term is added to the local training loss.}
\tim{In contrast, so far limited research~\cite{meerza2022fair,arasteh2023federated,kalabakov2024comparative,hsu2024cluster} has focused on speech-related FL techniques targeting dysarthric and elderly speakers, with most efforts directed toward Alzheimer's disease (AD) detection and very little toward speech recognition~\cite{hsu2024cluster}.}

\tim{Current FedAvg-based FL approaches face the following challenges when applied to dysarthric and elderly speech recognition: \textbf{1) data scarcity} further exacerbated by splitting the already limited dataset across clients; \textbf{2) data imbalance} where speakers with severe speech impairments or linguistic degradation have noticeably fewer words compared to those with milder impairments within a single client; and \textbf{3) speaker heterogeneity} among clients. Addressing these challenges necessitates the usage of effective, regularized federated learning techniques.}


\tim{To this end, this paper conducts a comprehensive exploration of regularized federated learning techniques to address the aforementioned challenges in advancing privacy-preserving dysarthric and elderly speech recognition. Specifically, we investigate regularization at different levels of the FL training process, including: \textbf{1) parameter-based regularization}, where a regularization term is added to the local training loss to align the parameters of the local model with those of the global model~\cite{MLSYS2020_1f5fe839}; \textbf{2) embedding-based regularization}, which, inspired by~\cite{Tun2023,GreidiCohen2024}, utilizes a regularization term to align intermediate embeddings of the local and global models during local training; and \textbf{3) novel loss-based regularization}, where local intermediate embeddings are passed through a frozen global model to generate pseudo-logits. These pseudo-logits are then aligned with local predictions using Kullback-Leibler (KL) divergence~\cite{kullback1951information}, which is further incorporated into the local training loss.} 
\tim{Performance evaluation is conducted on two benchmark healthcare datasets: \textbf{1)} UASpeech~\cite{kim2008dysarthric} dysarthric speech corpus; and \textbf{2)} DementiaBank Pitt~\cite{becker1994natural} for elderly speech, with FedAvg~\cite{pmlr-v54-mcmahan17a} serving as the base model aggregation strategy. In addition, the effects of applying regularization techniques at multiple positions, combining different regularization methods, and varying communication frequency are further analyzed.}

\tim{The main contributions of this paper are as follows:}

\tim{\textbf{1)} To the best of our knowledge, this paper presents the first systematic investigation of regularized FL techniques for privacy-preserving speech recognition in healthcare, with a focus on dysarthric and elderly populations. Different elements of the FL training process are explored, including parameter-based, embedding-based and the novel loss-based regularization. In contrast, prior research on regularized federated learning in speech recognition has been notably limited~\cite{nandury2021cross,pelikan2023federated}, and solutions addressing the challenges of FL-based dysarthric and elderly speech recognition have been rarely visited~\cite{hsu2024cluster}.}

\tim{\textbf{2)} Experiments on the benchmark UASpeech dysarthric speech and DementiaBank Pitt elderly speech corpora suggest that statistically significant word error rate (WER) reductions of up to 0.54\% absolute (1.69\% relative) and 0.55\% absolute (2.13\% relative) can be respectively obtained over the baseline FedAvg system without regularization. Combining the regularization techniques leads to further performance improvements.} 


\tim{The rest of the paper is organized as follows. Section~\ref{sec:std_fl} describes the standard FL based ASR systems. The three regularization techniques, i.e., parameter-, embedding- and loss-based regularization, are detailed in Section~\ref{sec:reg}. Section~\ref{sec:exp} presents experiments and
analysis on UASpeech and Dementiabank Pitt. Section~\ref{sec:conclusion} concludes the paper and discusses future work.}

\vspace{-0.3cm}
\section{\tim{Federated Learning Based ASR}}
\label{sec:std_fl}
\begin{figure}[htbp]
    \centering
    \includegraphics[scale=0.3]{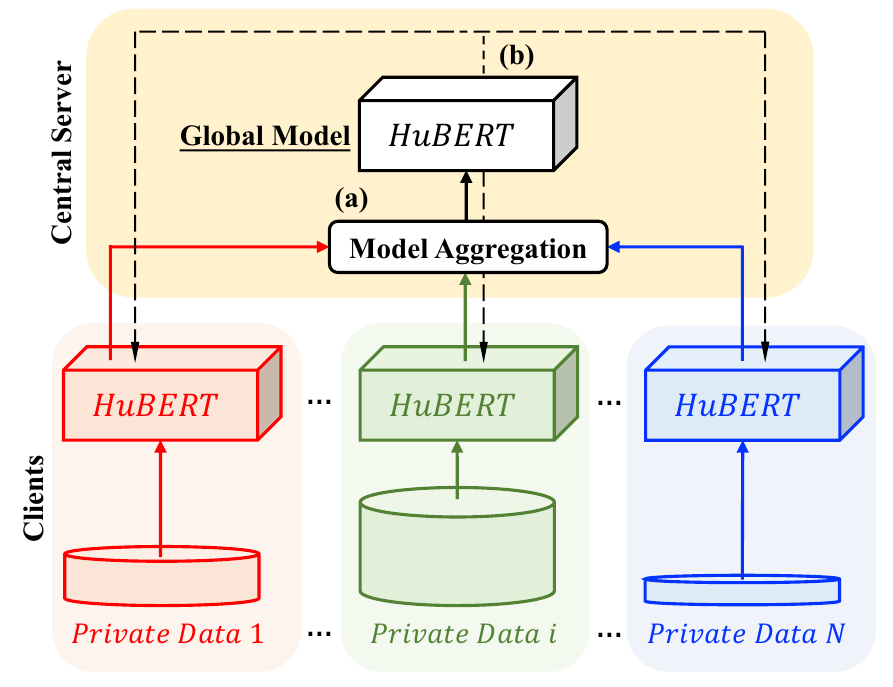}
    \caption{\tim{Illustration of federated learning based HuBERT~\cite{hsu2021hubert} ASR system. Following FedAvg~\cite{pmlr-v54-mcmahan17a}, each communication round involves: \textbf{(a)} aggregating the parameters of locally trained client-side models into a global model using a weighted sum, and \textbf{(b)} redistributing the global model to clients.}}
    \label{fig:FL_FedAvg}
\end{figure}

\tim{Unlike traditional distributed training~\cite{verbraeken2020survey} which first aggregates data from multiple sources and then redistributes the data among learners, federated learning (FL) preserves data locality to comply with privacy constraints. However, such a design inherently results in heterogeneous data distributions across clients. As illustrated in Fig.~\ref{fig:FL_FedAvg}, each client operates a local server for data storage and computation, interacting solely with a central server. During each communication round, only model parameters or gradients are exchanged between clients and the server. Local updates are sent to the central server, where they are aggregated into a global model (Fig.~\ref{fig:FL_FedAvg}(a)). The global model is then redistributed to clients as the starting point for client-side computation in the next round (Fig.~\ref{fig:FL_FedAvg}(b), dashed line).}

\vspace{-3pt}

\vspace{-5pt}

\tim{The widely used aggregation strategy for FL-based ASR is FedAvg~\cite{pmlr-v54-mcmahan17a}, which computes global parameters as a weighted sum of client parameters, with weights proportional to each client’s data fraction. Given the communication overhead~\cite{du2024communication}, synchronization is performed at carefully chosen frequency\footnote{\tim{The impact of communication frequency is analyzed in Section~\ref{sec:exp}.}}.}

\section{\tim{Regularized Federated Learning}}
\label{sec:reg}
\tim{Federated learning inherently introduces data heterogeneity among clients, which is further compounded by the challenges of data scarcity, data imbalance, and speaker diversity widely observed in dysarthric and elderly speech. To mitigate this, three regularization techniques at different levels are investigated, including parameter-based, embedding-based, and loss-based regularization. For model aggregation, we adopt FedAvg~\cite{pmlr-v54-mcmahan17a} as the default strategy.}

\vspace{-15pt}
\tim{\subsection{Parameter-based regularization}}
\label{sec:para_reg}

\begin{figure}[htbp]
    \centering
    \includegraphics[scale=0.2]{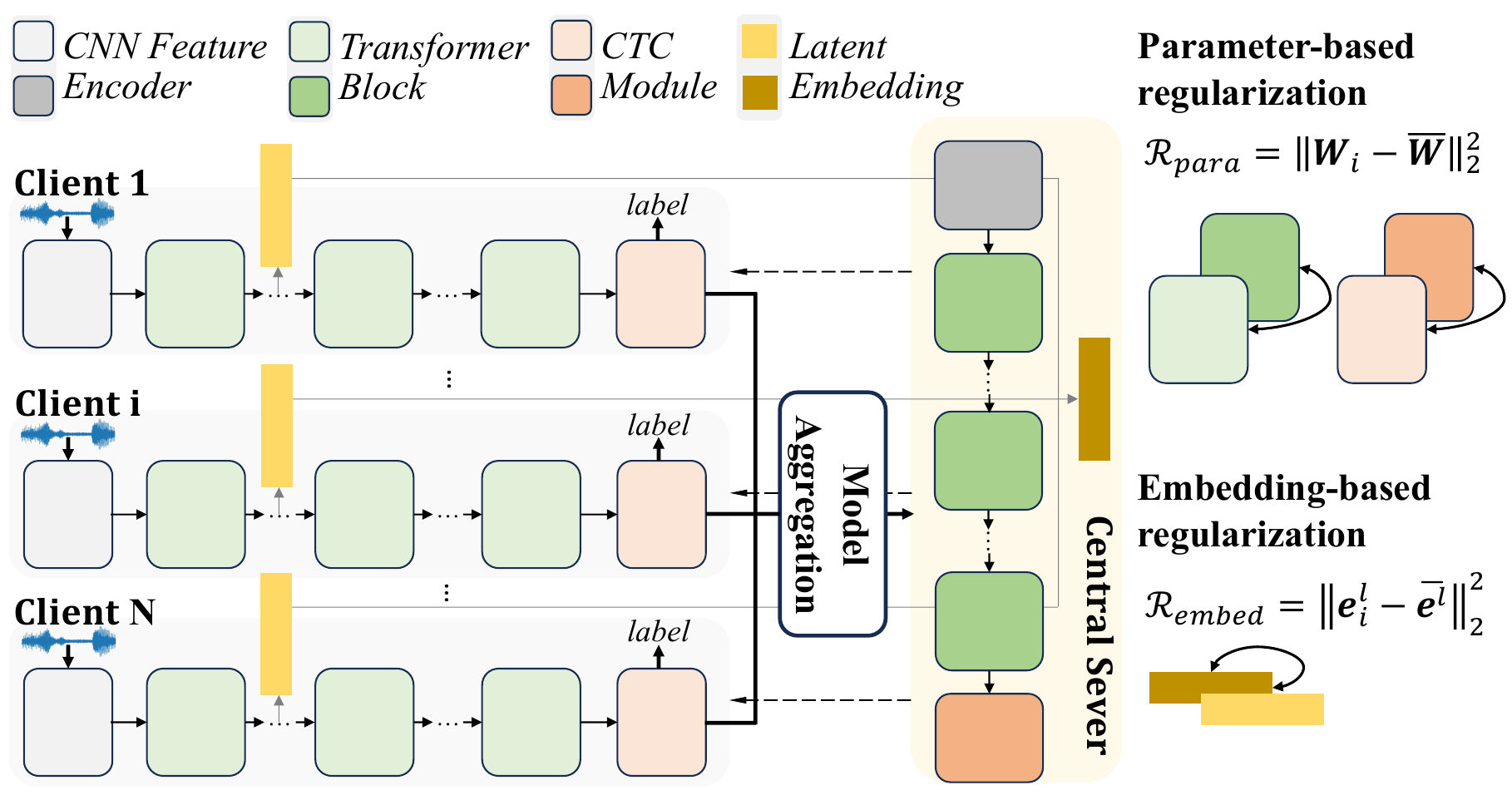}
    \caption{\tim{Schematic of \textbf{(a)} parameter-based regularization (upper right) and \textbf{(b)} embedding-based regularization (lower right) applied to the FedAvg-based HuBERT ASR system. Components with darker colors are on the central server side, while lighter-colored components are on the local client side.}}
    \label{fig:FL_reg_para_emb}
\end{figure}

\tim{Following FedProx~\cite{MLSYS2020_1f5fe839}, the global model parameters from the previous communication round are taken as a reference to regularize the local training process of the clients. As shown in the upper right of Fig.~\ref{fig:FL_reg_para_emb}, the squared L2 norm $\mathcal{R}_{para}$, which measures the Euclidean distance between the reference global model and the local model, is incorporated into the local training loss as a regularization term, given as:}

\vspace{-3mm}
\begin{equation}
    \label{eq:reg_para}
    \mathcal{R}_{para} = || \bm{W}_{i} - \overline{\bm{W}} ||_{2}^{2} 
\end{equation}


\noindent \tim{where $\mathbf{W}_i$ represents the local model parameters of the $i^{th}$ client at the current round, while $\overline{\mathbf{W}}$ denotes the global model parameters from the previous round.}


\vspace{-0.8cm}
\tim{\subsection{Embedding-based regularization}}
\label{sec:emb_reg}
\tim{Building on~\cite{Tun2023,GreidiCohen2024}, during each communication round, we extract latent embeddings from the local models, compute the average embedding over all utterances for each client, and then aggregate these averaged embeddings from all clients on the central server using a weighted sum\footnote{\tim{The weight is the client’s proportion of the total training data.}}. As demonstrated in the lower right of Fig.~\ref{fig:FL_reg_para_emb}, the aggregated embedding serves as a reference to constrain the local models' latent embeddings in the next round, where the local training loss integrates the squared L2 norm between these embeddings. This regularization term $\mathcal{R}_{embed}$ is given as:}

\begin{equation}
    \label{eq:reg_embed}
    \mathcal{R}_{embed} = || \bm{e}_{i}^{l} - \overline{\bm{e}_{i}^{l}} ||_{2}^{2} 
\end{equation}

\noindent \tim{where $\bm{e}_i^{l}$ represents the embedding of the $i^{th}$ client after the $l^{th}$ Transformer block, and $\overline{\bm{e}_i^{l}}$ denotes the aggregated embedding.}

\tim{\subsection{Loss-based regularization}}
\label{sec:loss_reg}

\begin{figure}[htbp]
    \centering
    \includegraphics[scale=0.3]{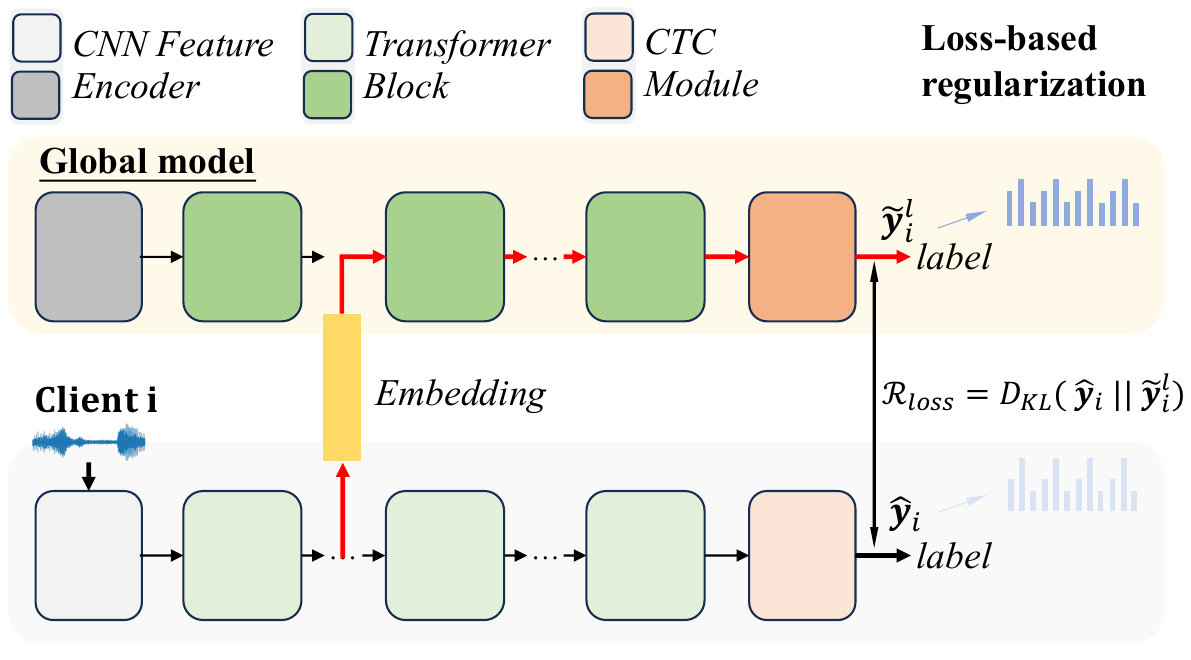}
    \caption{\tim{Schematic of the proposed loss-based regularization applied to the FedAvg-based HuBERT ASR system. For simplicity, one client is shown as an example. Darker-colored components represent the global model, while lighter-colored components correspond to the local model.}}
    \label{fig:FL_reg_loss}
\end{figure}

\tim{To further investigate regularization across different aspects of the training process, we propose a novel loss-based regularization approach. As illustrated in Fig.~\ref{fig:FL_reg_loss}, rather than directly constraining the latent embedding, we feed the local embedding of the $i^{th}$ client into the global model from the previous communication round and execute a feedforward pass to derive the output probability distribution (Fig.~\ref{fig:FL_reg_loss}, red bold line). This distribution is then leveraged to regularize the output distribution of the $i^{th}$ client by incorporating the Kullback-Leibler (KL) divergence~\cite{kullback1951information} between these two distributions into the local training loss. The regularization term $\mathcal{R}_{loss}$ is defined as:}

\begin{equation}
    \label{eq:reg_loss}
    \mathcal{R}_{loss} = D_{KL}(\hat{\bm{y}}_{i} \parallel \tilde{\bm{y}}_{i}^{l}) 
\end{equation}


\noindent \tim{where $D_{KL}(\cdot)$ denotes the KL divergence. $\hat{\bm{y}}_{i}$ represents the output distribution of the $i^{th}$ client, and $\tilde{\bm{y}}_{i}^{l}$ denotes the output distribution obtained by passing the $i^{th}$ client's embedding after the $l^{th}$ Transformer block through the global model.}

\tim{The embedding-based regularization and loss-based regularization can be applied at multiple positions in the model. Furthermore, as these three regularization techniques operate at different levels, their usage can be combined.}

\section{\tim{Experiments and Results}}
\label{sec:exp}

\subsection{\tim{Task description}}
\label{sec:task}

\tim{\textbf{The English UASpeech} corpus~\cite{kim2008dysarthric} is the largest publicly available and widely used dataset for dysarthric speech recognition. It comprises an isolated word recognition task with approximately 103 hours of speech data from 29 speakers, among whom 16 are dysarthric speakers and 13 are healthy control speakers. The dataset includes 155 common words and 300 uncommon words and is further divided into three blocks B1, B2 and B3. The same 155 common words are used across all blocks, while the 300 uncommon words differ between blocks. In our experiments, we focus \textbf{exclusively on the 16 dysarthric speakers} and exclude the healthy control speakers. B1 and B3 are used as for training, while B2 is used for evaluation. After removing silence, the training set and the test set contain 17.8 hours (52785 utterances) and 9 hours of audio (26520 utterances) in total, respectively. Speech intelligibility assessment is available for the 16 dysarthric speakers, divided into four groups: ``very low'' (VL), ``low'' (L), ``mid'' (M) and ``high'' (H).}



\noindent \tim{\textbf{The English DementiaBank Pitt}~\cite{becker1994natural} corpus is the most widely used publicly available dataset for speech-based diagnosis of Alzheimer’s Disease (AD).
It comprises 33 hours of cognitive impairment assessment interviews between 292 elderly participants and the associated clinical investigators. The training set includes 688 speakers (244 elderly participants, 444 investigators), while the development and evaluation sets\footnote{\tim{The evaluation set includes the 48 speakers' Cookie Theft recordings from the ADReSS challenge test set~\cite{luz2020alzheimer}, while the development set contains their recordings from other tasks, if available.}} respectively consist of 119 (43 elderly, 76 investigators) and 95 speakers (48 elderly, 47 investigators). There is no speaker overlap between the training set and either the development or evaluation sets. Silence stripping~\cite{ye2021development} produces a 15.7-hour training set (29682 utterances), a 2.5-hour development set (5103 utterances), and a 0.6-hour evaluation set (928 utterances).}


\subsection{\tim{Experiment setup}}
\label{sec:setup}


\tim{\textbf{Model configuration:} We adopt the state-of-the-art HuBERT~\cite{hsu2021hubert} model\footnote{\tim{https://huggingface.co/facebook/hubert-large-ls960-ft}} fine-tuned on 960 hours of Librispeech~\cite{panayotov2015librispeech} as the foundation for performing federated learning\footnote{\tim{https://apple.github.io/pfl-research/}} on dysarthric and elderly speech. The default model aggregation strategy is FedAvg~\cite{pmlr-v54-mcmahan17a}. A Connectionist Temporal Classification (CTC)~\cite{graves2006connectionist} model with a single fully connected layer is added on top of the CNN feature encoder and a stack of 24 Transformer blocks. In each communication round, the clients perform local training for one epoch, with the number of communication rounds set to 100. The penalty weights for parameter-based, embedding-based, and loss-based regularization are empirically set to 0.01, 0.001, and 0.01, respectively. Parameter-based regularization is applied to all model parameters except those of the CNN feature encoder, while embedding-based and loss-based regularization can be applied after the $6^{\text{th}}$, $12^{\text{th}}$, $18^{\text{th}}$, or $24^{\text{th}}$ Transformer block, or in any combination of these positions. Experiments are conducted using two Nvidia A40 GPUs. A matched pairs sentence-segment word error (MAPSSWE)~\cite{bisani2004bootstrap} based statistical significance test is performed with a significance level of $\alpha = 0.05$.}



\noindent \tim{\textbf{Client partitioning:} For the UASpeech dysarthric speech task, the training data of each of the 16 dysarthric speakers is assigned to a separate client, resulting in a total of 16 clients, each containing 0.71 to 1.54 hours of audio (1785 to 3570 utterances). For the DementiaBank Pitt elderly speech task, all conversations between each participant in the training set and the corresponding investigators are extracted, producing 244 conversation pairs in total. These 244 pairs are randomly divided into 10 non-overlapping sets, with each set assigned to one client, resulting in 10 clients with 1.35 to 1.75 hours of audio (2545 to 3437 utterances) per client.}



\vspace{-0.3cm}
\subsection{\tim{Result analysis}}
\label{sec:result}

\begin{table}[!htbp]
    \centering
    \caption{\tim{Performance of parameter (``para.''), embedding (``embed.'') and loss based regularization for federated learning on \textbf{UASpeech}. Here ``6'', ``12'' ``18'' and ``24'' refer to the the $6^{\text{th}}$, $12^{\text{th}}$, $18^{\text{th}}$, or $24^{\text{th}}$ Transformer block. ``VL'', ``L'', ``M'' and ``H'' are ``very low'', ``low'', ``mid'' and ``high'' speech intelligibility. $^{\dag}$ denote a statistically significant improvement $(\alpha = 0.05)$ obtained over the baseline FedAvg system (Sys.1).}}
    \label{tab:ua_reg_main}
    \setlength{\tabcolsep}{1pt} 
    \begin{tabularx}{\columnwidth}{c|c|Y|Y|Y|Y|cccc|c}
    \hline\hline
        \multirow{3}{*}{Sys.} & \multicolumn{5}{c|}{Regularization} & \multicolumn{5}{c}{UASpeech WER\%}  \\ 
    \cline{2-11}
         & \multirow{2}{*}{Method} & \multicolumn{4}{c|}{Position} & \multicolumn{4}{c|}{Speech Intelligibility} & \multirow{2}{*}{All} \\
    \cline{3-10}
       & & 6 & 12 & 18 & 24 & VL & L & M & H & \\   
    \hline\hline
        0 & \multicolumn{5}{c|}{centralized training} & 64.03 & 34.89 & 21.37 & 6.05 & 28.87  \\ 
    \hline
         \rowcolor{gray!25} 1 & \multicolumn{5}{c|}{\xmark} & 72.39 & 38.78 & 22.78 & 6.71 & 32.04 \\ 
    \hline
        2 & para. & \multicolumn{4}{c|}{-} &  72.22 & 38.56 & 22.67 & 6.44$^{\dag}$ & 31.83 \\ 
    \hline\hline
        3 & \multirow{6}{*}{embed.} & \checkmark & & & & 72.25 & 38.57 & 22.66 & 6.46$^{\dag}$ & 31.85 \\
        4 & & & \checkmark & & & 72.23 & 38.55 & 22.67 & 6.43$^{\dag}$ & 31.83 \\
        5 & & & & \checkmark & & 72.20 & 38.58 & 22.69 & 6.42$^{\dag}$ & 31.83 \\
        6 & & & & & \checkmark & 72.10 & 38.58 & 22.67 & 6.42$^{\dag}$ & 31.81 \\
    \cline{1-1}\cline{3-11}
        7 & & & & \checkmark & \checkmark & 71.95 & 38.50 & 22.65 & 6.40$^{\dag}$ & 31.70$^{\dag}$ \\
        8 & & \checkmark & \checkmark & \checkmark & \checkmark & 71.74$^{\dag}$ & 38.44$^{\dag}$ & 22.62 & 6.36$^{\dag}$ & 31.66$^{\dag}$ \\ 
    \hline\hline
        9 & \multirow{6}{*}{loss} & \checkmark & & & & 71.80 & 38.48 & 22.63 & 6.38$^{\dag}$ & 31.68$^{\dag}$ \\
        10 & & & \checkmark & & & 71.70$^{\dag}$ & 38.44$^{\dag}$ & 22.60 & 6.32$^{\dag}$ & 31.62$^{\dag}$ \\
        11 & & & & \checkmark & & 71.75$^{\dag}$ & 38.44$^{\dag}$ & 22.62 & 6.35$^{\dag}$ & 31.65$^{\dag}$ \\
        12 & & & & & \checkmark & 71.95 & 38.50 & 22.65 & 6.40$^{\dag}$ & 31.75 \\
    \cline{1-1}\cline{3-11}
        13 & & & \checkmark & \checkmark & & 71.62$^{\dag}$ & 38.30$^{\dag}$ & 22.60 & 6.25$^{\dag}$ & 31.54$^{\dag}$ \\
        14 & & \checkmark & \checkmark & \checkmark & \checkmark & 71.55$^{\dag}$ & 38.22$^{\dag}$ & 22.58 & 6.20$^{\dag}$ & 31.50$^{\dag}$ \\
    \hline\hline
        15 & \multicolumn{5}{c|}{para.+embed.+loss} & \textbf{71.50}$^{\dag}$ & \textbf{38.22}$^{\dag}$ & 22.50 & 6.19$^{\dag}$ & \textbf{31.45}$^{\dag}$ \\
    \hline\hline
    \end{tabularx}
    \vspace{-0.5cm}
\end{table}

\tim{\textbf{Experiments on dysarthric speech:} Table~\ref{tab:ua_reg_main} compares parameter-based, embedding-based and loss-based regularization for federated learning on UASpeech. Several trends can be observed: \textbf{1)} All three regularization techniques lead to performance improvements over the baseline FedAvg system without regularization (Sys.2-14 vs. Sys.1), while the proposed loss-based regularization produces better performances (Sys.9-14 vs. Sys.2-8). \textbf{2)} By applying embedding-based (Sys.7-8) or loss-based regularization (Sys.13-14) at different positions\footnote{\tim{Sys.7 combines the two best-performing positions among Sys.3-6, whereas Sys.13 integrates the two best positions from Sys.9-12.}}, with statistically significant overall WER reductions of up to 0.38\% abs. (1.19\% rel., Sys.8 vs. Sys.1) and 0.54\% abs. (1.69\% rel., Sys. 14 vs. Sys.1) over the baseline FedAvg system, respectively. \textbf{3)} Combining the three regularizations leads to further performance improvement\footnote{\tim{Weight is set as 0.1, 0.1, 1 for para., embed. and loss regularization.}}(Sys.15). \textbf{4)} Compared with centralized training (Sys.0), there remains a performance gap in federated learning, highlighting the challenges of FL-based dysarthric speech recognition due to its nature.}

\noindent \tim{\textbf{Experiments on elderly speech:} The performance of the three regularization techniques on DementiaBank Pitt is presented in Table~\ref{tab:dbank_reg_main}. Trends similar to those on dysarthric speech (Table~\ref{tab:ua_reg_main}) are observed, with statistically significant overall WER reductions of up to 0.43\% abs. (1.67\% rel.) and 0.55\% abs. (2.13\% rel.) obtained by embedding-based regularization (Sys.8) and loss-based regularization (Sys.14) over the baseline FedAvg system without regularization (Sys.1), respectively.}

\begin{table}[htbp]
    \centering
    \caption{\tim{Performance of parameter (``para.''), embedding (``embed.'') and loss based regularization for federated learning on \textbf{DementiaBank Pitt}.  ``Dev'' and ``Eval'' are the development and evaluation sets. ``PAR'' and ``INV'' are elderly participants and clinical investigators. $^{\dag}$ denote a statistically significant improvement $(\alpha = 0.05)$ obtained over the baseline FedAvg system (Sys.1). Other naming conventions follow Table~\ref{tab:ua_reg_main}.}}
    \label{tab:dbank_reg_main}
    \setlength{\tabcolsep}{1pt} 
    \begin{tabularx}{\columnwidth}{c|c|Y|Y|Y|Y|cc|cc|c}
    \hline\hline
        \multirow{3}{*}{Sys.} & \multicolumn{5}{c|}{Regularization} & \multicolumn{5}{c}{DementiaBank Pitt WER\%}  \\ 
    \cline{2-11}
         & \multirow{2}{*}{Method} & \multicolumn{4}{c|}{Position} & \multicolumn{2}{c|}{Dev} & \multicolumn{2}{c|}{Eval} & \multirow{2}{*}{All} \\
    \cline{3-10}
       & & 6 & 12 & 18 & 24 &  PAR & INV &  PAR & INV & \\   
    \hline\hline
        0 & \multicolumn{5}{c|}{centralized training} & 31.62 & 16.43 & 22.93 & 15.87 & 23.52 \\ 
    \hline
         \rowcolor{gray!25} 1 & \multicolumn{5}{c|}{\xmark} & 34.14 & 18.52 & 25.38 & 16.76 & 25.80 \\ 
    \hline
        2 & para. & \multicolumn{4}{c|}{-} & 33.84 & 18.28 & 25.15 & 16.30 & 25.54\\ 
    \hline\hline
        3 & \multirow{6}{*}{embed.} & \checkmark & & & & 33.90 & 18.33 & 25.16 & 16.30 & 25.58 \\
        4 & & & \checkmark & & & 33.86 & 18.26 & 25.15 & 16.28 & 25.54 \\
        5 & & & & \checkmark & & 33.80 & 18.24 & 25.14 & 16.28 & 25.52 \\
        6 & & & & & \checkmark & 33.82 & 18.23 & 25.12 & 16.25 & 25.51 \\
    \cline{1-1}\cline{3-11}
        7 & & & & \checkmark & \checkmark &33.70$^{\dag}$ & 18.20 & 25.05 & 16.25 & 25.44$^{\dag}$ \\
        8 & & \checkmark & \checkmark & \checkmark & \checkmark & 33.65$^{\dag}$ & 18.18 & 25.01 & 16.22 & 25.37$^{\dag}$ \\ 
    \hline\hline
        9 & \multirow{6}{*}{loss} & \checkmark & & & & 33.65$^{\dag}$ & 18.23 & 25.04 & 16.25 & 25.44$^{\dag}$ \\
        10 & & & \checkmark & & & 33.65$^{\dag}$ & 18.20 & 25.01 & 16.22 & 25.40$^{\dag}$ \\
        11 & & & & \checkmark & & 33.68$^{\dag}$ & 18.22 & 25.01 & 16.25 & 25.44$^{\dag}$ \\
        12 & & & & & \checkmark & 33.70$^{\dag}$ & 18.25 & 25.05 & 16.25 & 25.50 \\
    \cline{1-1}\cline{3-11}
        13 & & & \checkmark & \checkmark & & 33.60$^{\dag}$ & 18.10 & 25.00 & 16.15 & 25.32$^{\dag}$ \\
        14 & & \checkmark & \checkmark & \checkmark & \checkmark & 33.50$^{\dag}$ & 18.00 & 24.91 & 16.11 & 25.25$^{\dag}$ \\
    \hline\hline
        15 & \multicolumn{5}{c|}{para.+embed.+loss} & \textbf{33.45}$^{\dag}$ & 18.00 & \textbf{24.85}$^{\dag}$ & 16.06 & \textbf{25.21}$^{\dag}$ \\
    \hline\hline
    \end{tabularx}
    \vspace{-0.5cm}
\end{table}

\subsection{\tim{Impact of communication frequency}}

\begin{figure}[!htbp]
    \centering
    \includegraphics[scale=0.45]{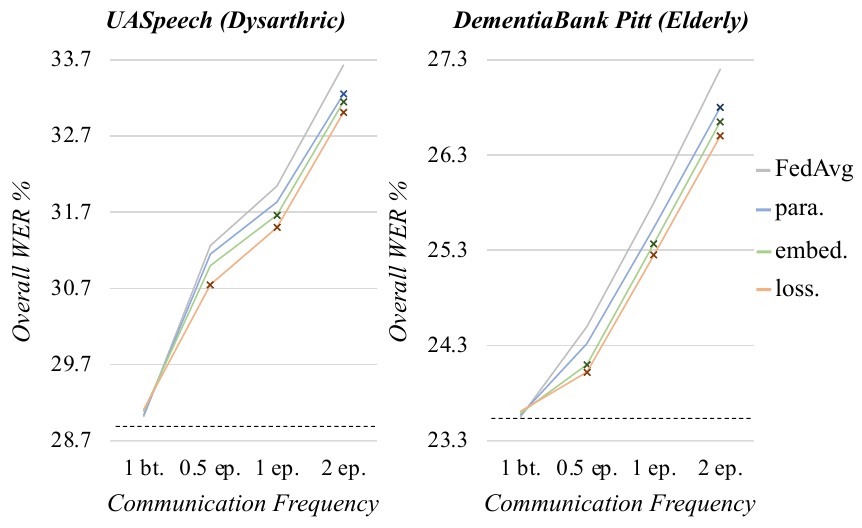}
    \caption{\tim{Impact of communication frequency on the performance (WER\%) of federated learning. ``bt.'' and ``ep.'' stand for batch and epoch. $\times$ denotes a stat. significant improvement $(\alpha = 0.05)$ over the baseline FedAvg system. The dashed line represents the performance of centralized learning (WER 28.87\% for UASpeech and 23.52\% for DementiaBank Pitt).}}
    \label{fig:communication}
\end{figure}

\tim{As communication overhead is a major challenge in federated learning~\cite{du2024communication}, we further investigate the impact of communication frequency on model performance. As shown in Fig.~\ref{fig:communication}, we vary the number of local updates performed by clients in each round, including an extreme case where communication occurs after every batch. To ensure a fair comparison, the total number of communication rounds is adjusted to keep the total training steps the same. Communicating after every batch approaches the performance of centralized learning (Fig.~\ref{fig:communication}, dashed line) on both datasets while keeping the data local, but this comes at the cost of 9 times increase in training time compared to the default setting of communicating once per epoch. Furthermore, the proposed loss-based regularization demonstrates a consistent and statistically significant performance improvement over the unregularized FedAvg systems (Fig.~\ref{fig:communication}, grey lines) across all communication frequencies, except in the 1-batch scenario.}

\review{On top of the communication costs incurred in the standard FedAvg training, the additional costs for these three regularization methods are: 1) 0 for parameter-based and loss-based embedding, as the former contrasts the global model with the local model while the latter feeds the local embedding to the global model for the loss comparison, both performed on the client side; 2) $O(d_{embed})$ for embedding-based regularization \timreview{for transmitting the embeddings}, where $d_{embed}$ is the size of the embeddings. When all three regularization techniques are combined, the additional costs are $O(d_{embed})$.}

\vspace{-0.4cm}

\vspace{-0.6cm}
\section{Conclusions}
\label{sec:conclusion}


\tim{This paper systematically investigates regularized FL techniques for privacy-preserving dysarthric and elderly speech recognition, i.e., parameter-, embedding-, and novel loss-based regularizations. Experiments on UASpeech and DementiaBank Pitt show that regularized FL systems consistently outperform FedAvg, while increasing communication frequency narrows the gap to centralized training. Future research will focus on speech-pattern driven regularization techniques.}

\vspace{-0.3cm}
\section{Acknowledgements}
This research is supported by Hong Kong RGC GRF grant No. 14200220, 14200021, 14200324 and Innovation Technology Fund grant No. ITS/218/21.

\bibliographystyle{IEEEtran}
\bibliography{new_bib}

\end{document}